\documentclass[s,pdf]{iucr}              

     \journalcode{S}              

\usepackage{graphicx}
\usepackage{textcomp}
\usepackage{epsfig}
\usepackage{array}
\usepackage{multirow}
\usepackage[utf8]{inputenc}

\begin{document}                  



\title{The Adaptive Gain Integrating Pixel Detector at the European XFEL}


\author[a]{Aschkan}{Allahgholi}
\author[a]{Julian}{Becker}
\author[a]{Annette}{Delfs}
\author[b]{Roberto}{Dinapoli}
\author[a]{Peter}{Goettlicher}
\author[b]{Dominic}{Greiffenberg}
\author[b]{Beat}{Henrich}
\author[a]{Helmut}{Hirsemann}
\author[a]{Manuela}{Kuhn}
\author[d]{Robert}{Klanner}
\author[a]{Alexander}{Klyuev}
\author[e]{Hans}{Krueger}
\author[a]{Sabine}{Lange}
\author[a]{Torsten}{Laurus}
\author[a]{Alessandro}{Marras}
\author[b]{Davide}{Mezza}
\author[b]{Aldo}{Mozzanica}
\author[a]{Magdalena}{Niemann}
\author[a]{Jennifer}{Poehlsen}
\author[d]{Joern}{Schwandt}
\author[a]{Igor}{Sheviakov}
\author[b]{Xintian}{Shi}
\author[a]{Sergej}{Smoljanin}
\author[a]{Lothar}{Steffen}
\author[c]{Jolanta}{Sztuk-Dambietz}
\author[a]{Ulrich}{Trunk}
\author[]{Qingqing}{Xia}
\author[a]{Mourad}{Zeribi}
\author[b]{Jiaguo}{Zhang}
\author[a]{Manfred}{Zimmer}
\author[b]{Bernd}{Schmidt}
\cauthor[a,f]{Heinz}{Graafsma}{Heinz.Graafsma@desy.de}{}

\aff[a]{Deutsches Elektronen Synchrotron, Notkestrasse 85, 22607 Hamburg, \country{Germany}}
\aff[b]{Paul Scherrer Institut, Forschungsstrasse 111, 5232 Villigen, \country{Switzerland}}
\aff[c]{European XFEL, Holzkoppel 4, 22869 Schenefeld, \country{Germany}}
\aff[d]{University of Hamburg, Luruper Chaussee 149, 22761 Hamburg, \country{Germany}}
\aff[e]{University of Bonn, Nussallee 12, 53115 Bonn, \country{Germany}}
\aff[f]{Mid Sweden University, Holmgatan 10, S-85170 Sundsvall, \country{Sweden}}


\shortauthor{A. Allahgholi et. al.}







\maketitle                        


\begin{abstract}
The Adaptive Gain Integrating Pixel Detector (AGIPD) is an x-ray imager, custom designed for the European x-ray Free-Electron Laser (XFEL). It is a fast, low noise integrating detector, with an adaptive gain amplifier per pixel. This has an equivalent noise of less than 1~keV when detecting single photons and, when switched into another gain state, a dynamic range of more than 10$^4$ photons of 12~keV. In burst mode the system is able to store 352 images while running at up to 6.5~MHz, which is compatible with the 4.5~MHz frame rate at the European XFEL. The AGIPD system was installed and commissioned in August 2017, and successfully used for the first experiments at the Single Particles, Clusters and Biomolecules (SPB) experimental station at the European XFEL since September 2017. This paper describes the principal components and performance parameters of the system.
\end{abstract}


\section{Introduction}

With the start of the European XFEL, a new milestone is set in the field of x-ray research and many related fields due to the high coherence, pulse intensity and repetition rate of the x-ray pulses available at this facility. The superconducting accelerator provides up to 600~\textmu s long trains with up to 2700 pulses followed by an inter-train gap of 99.4~ms. Inside each train, consecutive pulses of typically less than 100~fs duration are spaced approximately 220~ns apart. This corresponds to an effective repetition rate of 4.5~MHz during a train. Each pulse contains up to 10$^{12}$ photons \cite{Altarelli}, which in many cases is sufficient to produce a complete scattering pattern from the sample with a single pulse. This means that the area detectors at the European XFEL not only have to be compatible with the high repetition rate of the source, but also need to have a dynamic range from single photons to 10$^4$ photons/pixel/pulse. More detailed and complete requirements can be found in \cite{Graafsma}. Dedicated detector development programs for the European XFEL were started more than 10 years before its inauguration, as it was clear that no existing detector would be able to meet the requirements imposed by the new facility. The Adaptive Gain Integrating Pixel Detector (AGIPD) was designed to fulfill as many of the general requirements as possible with a focus on the requirements most important for scattering experiments in the energy range from 7 to 15~keV, for which is has been successfully used in first user experiments \cite{Ilme} and \cite{SFX}. There has been interest to use the system at other bright sources to study fast dynamics down to the microsecond scale.

The AGIPD is not the only camera developed for use at the European XFEL. The LPD system \cite{Hart2012} is currently in use at the FXE beamline and the DSSC system \cite{Porro_2012} will be available soon. Other FELs are using other custom developed camera systems like the CSPAD \cite{Philipp2011} and ePIX detectors \cite{Blaj2016} at LCLS or the JUNGFRAU detector \cite{Redford2016} at the SwissFEL.

\section{System layout}

\begin{figure}
\includegraphics[width=\textwidth]{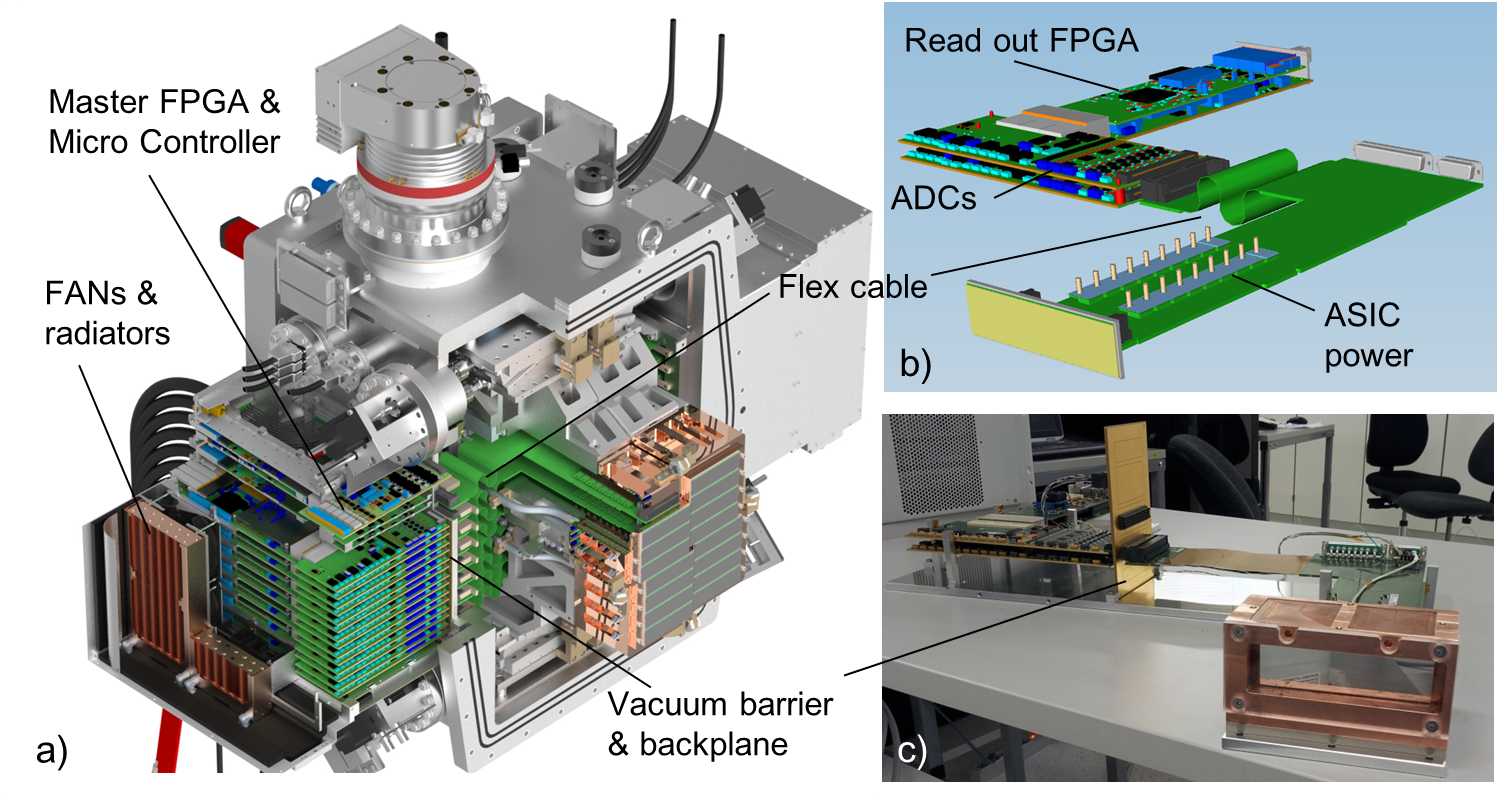}
\caption{a) CAD design of the AGIPD 1 million pixel detector with cuts to expose the arrangement of the electronics inside and outside of the vacuum vessel. b) CAD model of the electronics of a single tile. c) Picture of a single tile using a 2 port version of the vacuum backplane board.}
\label{fig1}
\end{figure}

The AGIPD camera consists of four individually moveable quadrants, each having four detector tiles with 512 x 128 pixels per tile, giving a total of 1024 x 1024, or roughly 1 million pixels. Fig. \ref{fig1}a shows a CAD design of the AGIPD 1 million pixel detector with cuts to expose the arrangement of the electronics inside and outside of the vacuum vessel.

Each detector tile consists of a front end module (FEM), an in-vacuum board that provides power to the FEM and routes signals, two ADC boards and a control and data IO board. Fig. \ref{fig1}b and c show a CAD drawing and a photograph of these boards, respectively.

Being glued to a ceramic board, each FEM consists of a monolithic pixelated silicon sensor responsible for absorbing the x-ray photons and creating an electrical signal per pixel which proportional to the sum of the energies of all simultaneously absorbed photons in that pixel. The silicon sensor is bump-bonded to 2~x~8 pixelated Application Specific Integrated Circuits (ASICs), which are responsible for signal integration and intermediate image storage. 

The front-end modules protrude into the attached sample interaction chamber and can be cooled to stabilize their temperature and improve performance.  Although there are many components in the vacuum chamber, vacuum levels better than $10^{-5}$~mbar have been reached in our labs. Note that this was without attaching the system to the sample interaction chambers at the European XFEL, so the vacuum levels obtained during XFEL experiments might be different.

The detector logically and electrically divides into two wings. Each wing consists of the ADC boards and control and data IO board of each tile (one set each per tile, 8 tiles per wing) as well as a vacuum backplane board, which acts as a vacuum barrier and routes signals in and out of the vacuum vessel, a micro controller board for slow control and a master FPGA board. These boards are located outside the vacuum chamber\footnote{The vacuum backplane forms a vacuum interface.} in a thermally sealed, water cooled housing. 

The two master FPGA boards, one for each side, provide the interface to the clock and control system, and control the detector tiles, including the FEMs. In the following sections the individual components and operational concepts are described in more detail.

\section{Mechanics}

\begin{figure}
\includegraphics[width=\textwidth]{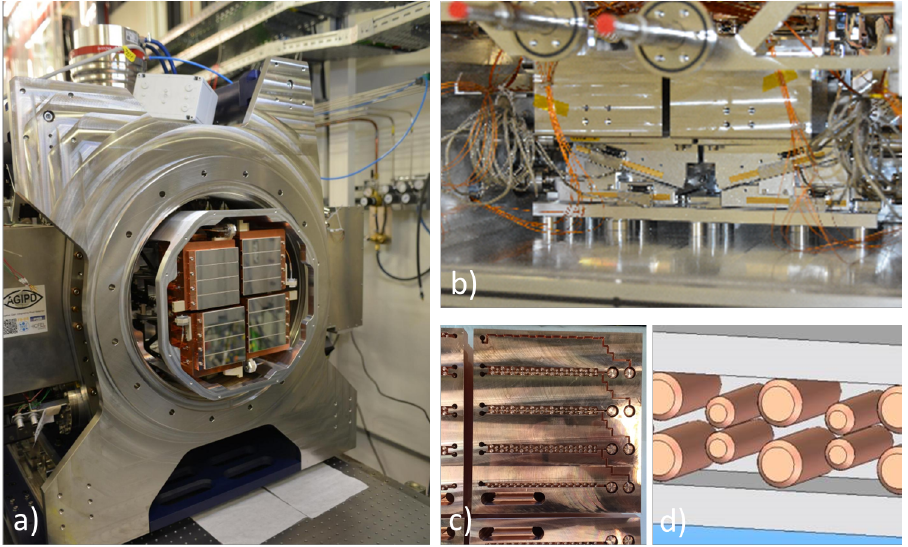}
\caption{a) Photograph of the AGIPD 1 million pixel system at the SPB beamline before mounting to the experimental chamber. The sensitive area is split into four independently movable quadrants. b) Wedge system of the bottom quadrants. c) Cooling channels of a single cooling block before electroforming and milling connector feedthroughs. d) 3D image of the pins inserted into the cooling channels to enhance the turbulence of the  flow}
\label{fig2}
\end{figure}

Fig. \ref{fig2}a shows the front view of the AGIPD 1 million pixel system installed at the SPB station of the European XFEL. The four independently movable quadrants allow the formation of a horizontal slit or a rectangular central hole with user selectable size for the direct beam to pass through. The movement is realized by mounting each cooling block on a motion stage formed by two wedges (Fig. \ref{fig2}b).

The cooling blocks were made by a combination of milling and electroforming techniques. In the first step, the basic shape of the cooling block was milled out of a solid block of HCP copper. This basic shape omitted details that would be included in the final milling step, but already included the cooling channels. To enhance the turbulence of the flow of the silicone oil coolant, holes were drilled into the channels and pins were inserted into the holes. Afterwards, the channels were covered by copper using an electroforming process. Initially omitted details, like connector feedthroughs, were defined in a final milling run, which also ensured the overall dimensions and tolerances of the cooling block. Fig. \ref{fig2}c and d show the channels with inserted pins before the cover deposition.

\section{Front end modules}

An AGIPD 1 million pixel detector incorporates 16 front end modules (FEMs). Each front end module uses a bump bonded hybrid of a monolithic silicon sensor with 128 x 512 pixels and 2 x 8 AGIPD ASICs. The power and signal contacts of the ASICs are wire bonded to gold plated pads on an LTCC (Low Temperature Co-fired Ceramics) board. The bias voltage of the sensor is provided via wire bond connections between sensor and LTCC on all four corners of the assembly. A PT1000 temperature sensor is located on the backside of the LTCC, next to a Samtec 500 pin connector, which connects the FEM to the interface electronics. The hybrid assembly is glued to a silicon heat spreader, which reduces thermal gradients across the tile. For the bonding between heat spreader and LTCC an adhesive with high thermal conductance is used that is able to handle the different coefficients of thermal expansion of silicon and ceramics. 

\begin{figure}
\includegraphics[width=\textwidth]{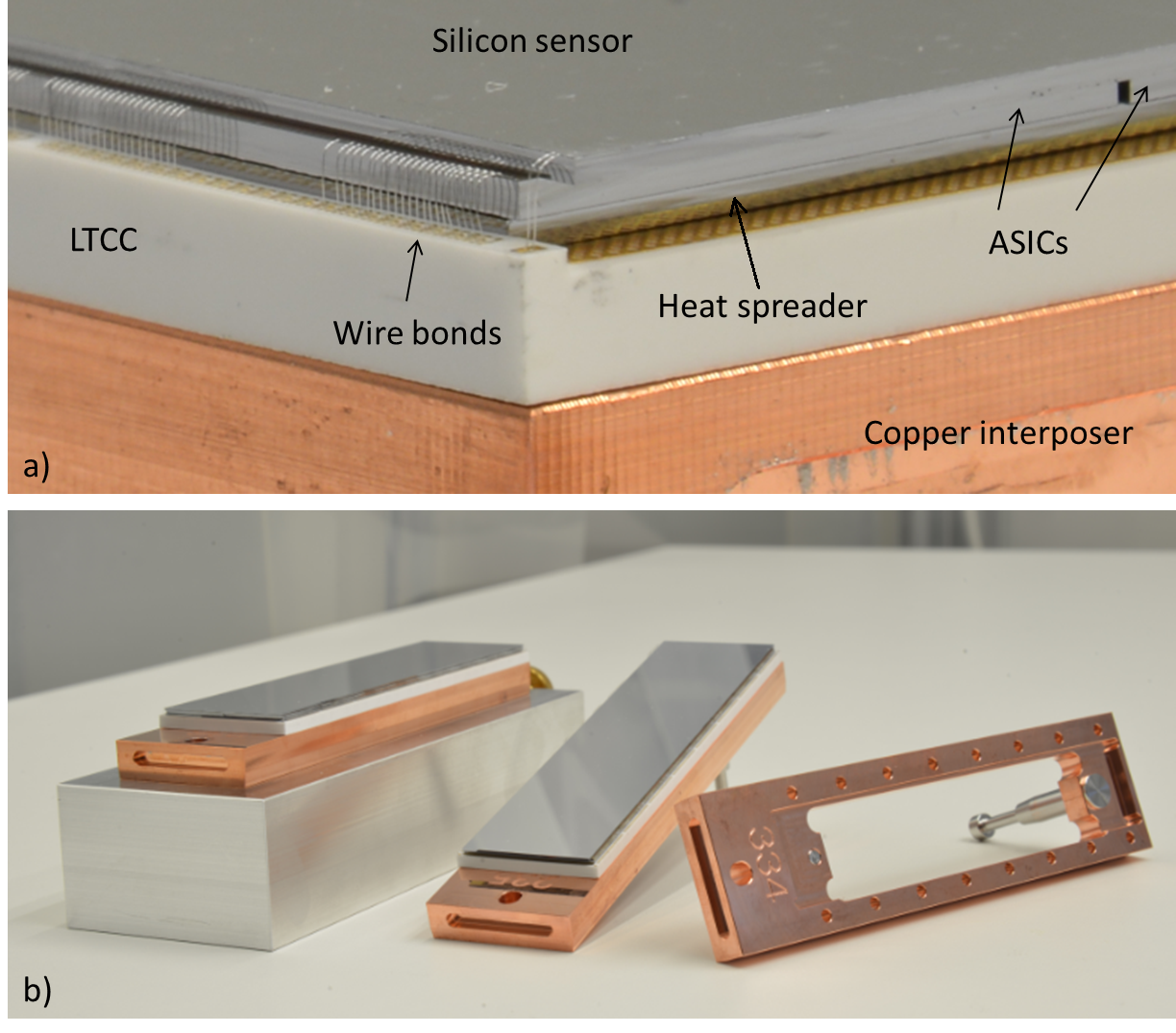}
\caption{a) Annotated macro photograph of the edge of a front end module, b) FEMs and copper interposers for handling and mounting.}
\label{fig3}
\end{figure}

For mounting and handling, the wire bonded assembly is screwed onto an interposer made from a copper alloy which provides sufficient rigidity and heat conductivity. The interposer features an insertion pin for defined mounting to the cooling block. To reduce the thermal resistance between the layers, the interface between LTCC and interposer is filled with vacuum compatible liquid gap filler and the interface between interposer and cooling block with graphite. Fig. \ref{fig3}a shows a macro photograph of the edge of a FEM, detailing sensor, ASICs, heat spreader, LTCC and interposer. Fig. \ref{fig3}b shows two fully assembled FEMs and a bare interposer.

In our lab the system has been tested at temperatures as low as -20~C on the LTCC, but the system can run at temperatures above that, which was also the case for early user experiments at the SPB beamline. The system is also compatible with operation at ambient pressure and temperature.

\subsection{The ASIC}

\begin{figure}
\includegraphics[width=\textwidth]{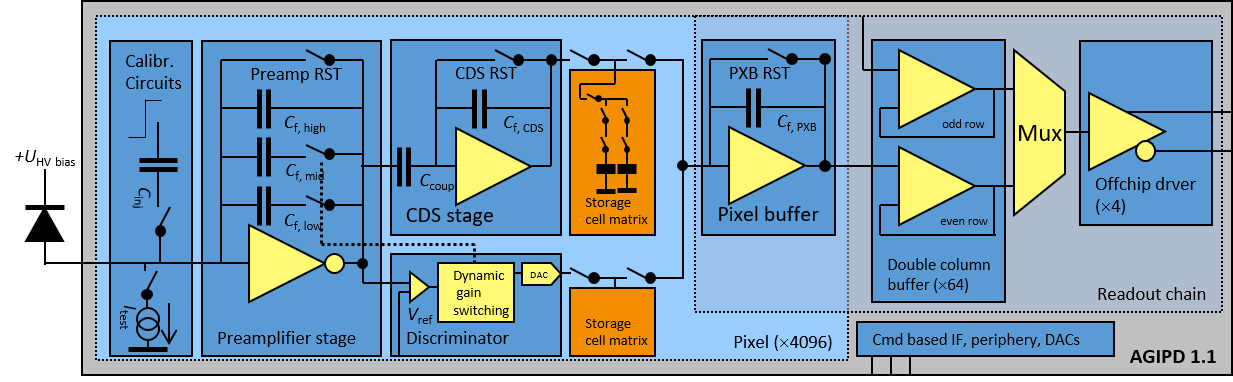}
\caption{Schematics of the AGIPD 1.1 readout ASIC.}
\label{fig4}
\end{figure}

The AGIPD 1.1 ASIC \cite{XS_NIM10, mezza1} forms the heart of the system; each ASIC incorporates 64 x 64 pixels and the necessary readout and control circuitry. It is manufactured in 130~nm CMOS technology, using radiation hardened layout techniques in most parts of the circuitry.

Fig. \ref{fig4} shows a schematic diagram of the pixel circuitry: the input of each pixel is formed by a resettable charge sensitive preamplifier, built around an inverter core. Its output feeds a discriminator and a correlated double-sampling (CDS) stage with two globally adjustable gain settings. Once the discriminator output exceeds the globally defined threshold, additional feedback capacitors of 3 or 10~pF are added to the 60~fF of the un-switched preamplifier feedback loop. This way the sensitivity of the preamplifier is adaptively decreased and the dynamic range is extended in two steps \cite{AGIPD_Shi}, where each pixel automatically and independently adapts its gain to the incoming signal. The CDS stage is used to remove noise from the reset switch and to suppress low frequency noise components \cite{Buttler}. 

The pixel response is recorded to an in-pixel memory at high speed, compatible with the 4.5 MHz requirement for the European XFEL \cite{UT_IEEE11}. Each memory address contains two separate pieces of information: a) the output voltage of the CDS stage, which is proportional to the detected signal, is stored on a 200 fF capacitor and b) the gain state of the pixel, encoded as a voltage, is stored on a 30 fF capacitor. 

The memory matrix occupies about 80\% of the pixel area and can store up to 352 images consisting of signal and gain information. The pixel size of (200~\textmu m)$^2$ is a compromise between resolution, analog performance and number of memory cells. 

The memory can be randomly accessed, providing the option of overwriting images or frame selective readout. At the European XFEL it is used to implement a veto system \cite{candc}. 

During readout of the chip another charge sensitive amplifier in each pixel is used to read the memory, which happens in parallel for each row of pixels. The further readout uses two interleaved column buses and four multiplexers, each serializing data from a block of 16 x 64 pixels. This parallelizing onto 4 outputs reduces the power consumption of the readout circuitry by reducing its speed. Instrumentation amplifiers convert the signal to differential levels, which are driven off-chip for subsequent digitization. 

A command based control circuit provides all the signals for memory access, read and write operations to the pixels. It uses a 3 line serial current mode logic interface and also provides slow control tasks, like the programming of internal timings and on-chip biases generated by digital to analog converters \cite{mezza2}.

\subsection{The Sensor}

Much like the ASIC, the sensor design was driven by the requirements set forth by the expected experiments and performance of the European XFEL \cite{Graafsma}. The main requirements were a thickness of 500~\textmu m in order to reach sufficiently high quantum efficiency and a tolerance of a total dose of 1~GGy during the expected lifetime of the detector. All of this should be accomplished while making large area sensors that allow minimizing the dead area of the final detector system.

In addition to the challenge of radiation tolerance, the impact of plasma effects due to the high number of instantaneously absorbed photons was investigated \cite{plasma}. When many photons are locally absorbed, sufficient electron-hole pairs are created to form a plasma. This plasma shields its core from the electric field required to drift the charges to the readout ASICs. 

Diffusion will eventually expand the plasma and lower its density enough for the drift field to take over, but this process takes time. As a result the charge cloud will spread laterally, potentially degrading the spatial resolution, and the total charge collection time will increase, potentially piling up with the next photon pulse. 

These studies concluded that the sensor should be constructed using p$^+$ electrodes in a highly resistive n-type bulk, thus collecting holes. Also a voltage of at least 500~V should be applied to the AGIPD sensor to suppress the consequences of the plasma effects as much as possible. At this voltage the time to collect at least 95\% of the deposited charge is less than 60~ns for tightly focused spots (3~\textmu m rms) of up to 10$^5$ 12~keV photons.

Surface damages are the dominant type of radiation damage for the sensor, as the damage threshold for silicon bulk damage is far above 12 keV, the original design photon energy of the European XFEL. 

Therefore, surface damages, namely the creation of positive charges in the SiO$_2$, and the introduction of traps at the Si-SiO$_2$ interface, were studied in detail with numerous irradiation campaigns \cite{Klanner} to establish the relevant parameters, which in turn were used for sensor optimization studies \cite{Schwandt1}. 

These studies showed that several design parameters, i.e. oxide thickness, pixel implant depth, and metal overhang, have significant influence on the ability to operate the sensor at a high voltage. Some of the findings show conflicting results for the situation before and after irradiation, i.e., a thick oxide is preferred before irradiation and a thin oxide after irradiation \cite{Schwandt_2013}. 

Finally, a compromise layout was chosen that fulfilled the design specifications and showed a breakdown voltage above 800~V in simulation \cite{Schwandt2}. This layout was produced and its radiation tolerance was studied \cite{Zhang}, finding sufficient performance for operation at the European XFEL.

The dead area of the final system is minimized for each front end module by using a monolithic sensor bump bonded to 2~x~8 individual ASICs. The large sensor guarantees that there are no blind spots within the sensitive surface (0\% dead area), however since the ASICs cannot physically touch each other the pixels horizontally in between two ASICs are twice as wide (400~\textmu m x 200~\textmu m), covering the area of two ordinary pixels. Pixels vertically in between two ASICs have normal size. 

The entire sensitive area is surrounded by a guard ring that is 1.2 mm (or 6 pixels) wide. A detailed description of the sensitive and non-sensitive areas and the implications for coherent diffraction imaging at the SPB instrument of the European XFEL can be found in \cite{Klaus}. Disregarding intentional gap between quadrants caused by the moving apparatus each quadrant has less than 15\% insensitive area, which increases to less than 18\% for the whole system, still disregarding intentional gaps.

\section{Signal handling, data preparation and control}

As indicated earlier, the AGIPD 1 million pixel detector is made of two electronically independent halves. Each half consists of a master FPGA board, a micro controller based slow control unit, an 8 port vacuum barrier board and eight tiles consisting of front end modules and their readout boards. 

The 8-port vacuum barrier board, which interconnects between the different boards, is realized using a multi-layer printed circuit board which acts as a vacuum barrier and as a backplane for signal distribution. 

The master FPGA acts as an interface device between the detector, the clock and control system of the European XFEL \cite{candc} and the control PC. It receives configuration, ’start’ and ’stop’ signals from the control computer; receives bunch synchronized clocks and classification flags; synchronizes the operation of the ASICs and fast ADCs; and triggers the read out FPGAs of the tiles. 

The boards constituting the readout electronics of each tile are located either inside the vacuum vessel or in the enclosed air environment of the external housing. 

From each front end module 64 differential analog signal lines are guided via the in-vacuum board and vacuum barrier board to the analog boards. The flexible part of the in-vacuum board (indicated in Fig. \ref{fig1}a and b) allows the movement of the quadrants. 

The ADCs (Analog Devices AD9257) sample at a resolution of 14~bit and operate at a frequency of up to 33~MHz, which results in a minimum possible read out time of approximately 22~ms for 352 frames (signal amplitude and gain state information are read separately, no overheads), well within the 99.4~ms inter train spacing at the European XFEL. At 14~bit resolution the ADC noise and the quantization errors do not contribute significantly to the overall noise of the system.

Each readout FPGA \cite{Sheviakov,Xia} orders the information of the 64 ADC channels of its tile into a single frame and sends the data via a 10~GbE optical link to the data acquisition system of the European XFEL. 

The slow control board monitors the status of the detector by collecting data on supply currents and voltages, as well as temperature, humidity and cooling fan information. 

During the detector start-up phase the experimental control computer sends commands to the micro controller to power the electronics sequentially. Collected monitor and status information can be queried by the slow control computer of the experiment which adds time stamps and makes the data available to the XFEL system. 

In addition, the slow control board serves as a second level interlock, transmitting hard wired flags to the experimental programmable logic controller (PLC) to initiate a shutdown, if operational conditions are outside pre-defined safety margins. The architecture of the electronics is discussed in more detail in \cite{goettlicher}.

\section{Calibration}

The performance of the AGIPD ASIC has been extensively tested and documented over the years \cite{mezza2,mezza1}. Since every pixel can be in any of its three gain states in any image (i.e., memory cell) the current calibration procedure calibrates each memory cell individually for all three gain settings requiring more than 2.8 billion calibration parameters for a 1 million pixel detector system.

In a big system of one million pixels or more, calibrating all relevant parameters using only external sources is not feasible in a time efficient manner. Therefore the ASIC includes two internal calibration sources - a current source and a pulsed capacitor - to enable dynamic range scans without the need for external stimulus. Both sources combine elements that are global to the chip and local in each pixel. The current source is implemented as a distributed current mirror and each pixel features a calibration capacitor while the voltage step is generated in the periphery.

\begin{figure}
\includegraphics[width=\textwidth]{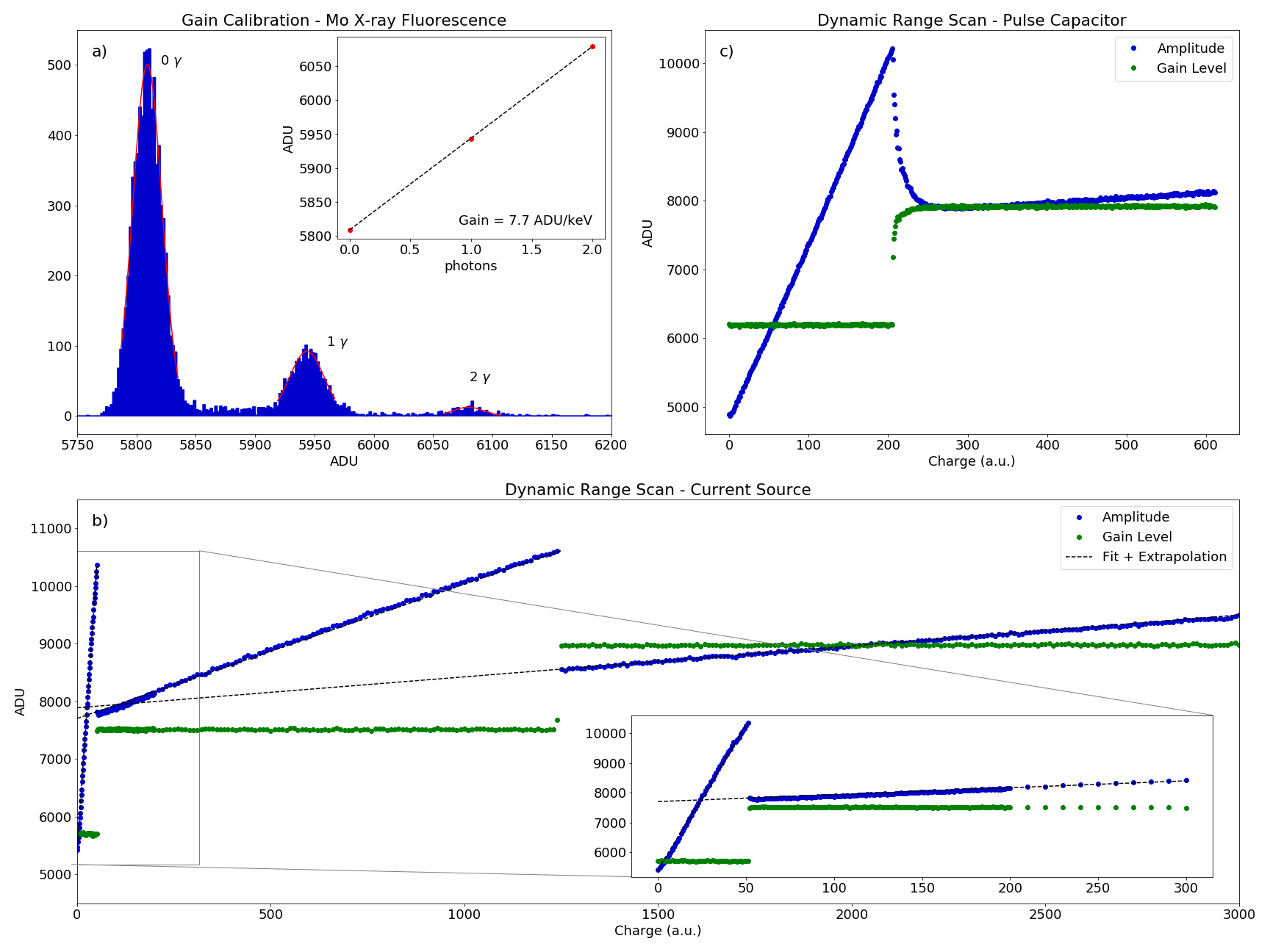}
\caption{a) Histogram of 10,000 frames for a single memory cell of a  single pixel illuminated with characteristic x-rays from molybdenum. The gain factor is derived from the mean peak-to-peak distance. b) Using the internal current source all three gain levels can be sampled. c) The pulsed capacitor samples only the high gain state and parts of the medium gain state. The non-ideal transition between the two gains is caused by the finite bandwidth of the calibration circuit.}
\label{fig5}
\end{figure}

Each data point from the detector contains two values per pixel per memory cell: the analog signal value and the encoded gain state information. Each value is expressed as Analog to Digital converter Units (ADU) in the raw data stream of the detector.

The internal sources allow the determination of valuable information. For the analog signal value, that is the relative gain of all 3 states within each pixel and the relative gain of the pixels with respect to each other. For the encoded gain state used, that is the discrimination threshold that indicates the different states of the gain.

The offsets for each gain state are measured without illumination (dark frames) and the gain of the high gain state is measured for each pixel using a flood illumination with characteristic x-rays. The characteristic x-rays were generated by illuminating a metal foil with an x-ray tube. The distance between foil and sensor was approximately 10-20~cm, depending on the foil, to ensure sufficient count rate in all pixels without creating a strong gradient in count rate over the module.

In a last step the calibration data are matched and merged, resulting in independent calibration constants for each memory cell in each pixel. Each memory cell is characterized by 8 parameters: 3 offsets (in ADU, one per state), 3 gain values (in ADU/keV, one per state), and 2 thresholds (in ADU) for state discrimination. This totals more than 2.8 billion calibration parameters in a 1 million pixel detector system.

Figure \ref{fig5} shows examples of intermediate results during the calibration procedure. Fig. \ref{fig5}a shows the histogram of 10,000 analog signal values measured for a single memory cell of a single pixel. The data were acquired using illumination with characteristic x-rays from a molybdenum foil. The absolute gain of this memory cell of this pixel is given by the distance between the peaks in the histogram (noise to single photon or photon to photon). The average gain value of a typical FEM is 7.7 ADU/keV +- 3.4 \%. 

The integration time during data taking was increased to 50~\textmu s, a value significantly beyond the 130~ns typically used for experiments at the European XFEL. This was done to increase the number of detected photons per frame to at least 1000 and thereby reduce the statistical uncertainty of the fit. Due to the long integration time the noise of the system (width of the peaks) is much higher than for typical experimental conditions. The noise measurement of the high gain state is extracted from dark frames taken under experimental conditions (not shown) and typical values for the Equivalent Noise Charge (ENC) are 320 and 240 electrons in CDS gain low or high, respectively. For the other gain states the noise is higher, but still below the Poisson limit \cite{mezza1}.

The internal current sources allow extrapolation of the absolute calibration of the high gain state to other states. The procedure using the internal current source utilizes injecting a constant current to the input of the pre-amplifier while sweeping the integration time. 

An example of a current source scan is shown in Fig. \ref{fig5}b. From linear fits to the data the ratios of low, medium and high gain are determined\footnote{The deviations from the ideal behavior at the beginning and end of each gain state are excluded from this fit.}. The high- and medium gain regime can also be scanned with a pulsed capacitor as shown in Fig. \ref{fig5}c. 

The pulsed capacitor scans the dynamic range by gradually increasing the height of the applied voltage step while keeping the integration time constant. The encoded gain state values and the corresponding discrimination thresholds can be extracted from both current source and pulsed capacitor scans, and are shown by the green dots in Fig. \ref{fig5}b and c.

Recording the required data to extract the calibration data typically takes 12-14 hours and occupies more than 20~TB of disk space. The largest contribution to this is the current source scan, which typically takes 10-11 hours and generates roughly half of the data volume. The smallest contribution to this has the x-ray illumination, which is the only time the detector must be illuminated during the calibration data taking process, with about 30 min duration and 0.4~TB of data volume. 

Analysis of the data on the DESY MAXWELL cluster using routines that have been optimized for parallel data processing can be performed in less than a day.

The electrical calibration described so far is essential for every experiment. In addition to this, the mechanical calibration, which determines the position of each pixel in all three dimensions and especially in relation to the beam, is of great importance for many experiments as well. 

The detector mechanics is designed and built with high precision, but the movability of the four quadrants requires a quick, robust and accurate method of calibration for the position and orientation of the FEMs at any time. On top of the uncertainties introduced by the movements, small displacements and tilts of the FEMs\footnote{Since each FEM uses a monolithic silicon sensor which is defined by photolithography to precisions much better than 1~\textmu m the displacements of the pixels in each module w.r.t. their ideal positions in the pixel matrix is negligible compared to the other displacements described here.} with respect to ideal positions are unavoidable as their manufacture involves gluing steps which are naturally limited in their positioning accuracy. 

A commonly used approach to calibrate the absolute position of each FEM in space is to take diffraction data of a well-known sample and fit the detector geometry \cite{crystfel}.

The detector has just entered routine operation at the European XFEL and first crystal structures have been successfully determined \cite{Ilme,SFX}. For these experiments the detector was calibrated using the procedures described here. However the procedures and recommended intervals are likely to evolve as the system gets used more often. 

Currently, we recommend to take dark data for offset determination before and after each practical block of scientific data taking to account for any small drifts, e.g., of the temperature, that might occur during the scientific data taking process. Rechecking of the absolute gain and the gain ratios between the gain states using x-rays and the internal sources should be done periodically, as these might change over longer periods of time, e.g., due to accumulation of radiation damage. 

While the system has been designed to be radiation hard, as detailed in earlier sections, forecasting the point at which radiation damage effects do show up is currently not possible for us.

Further, we recommend an electrical recalibration of the detector every time there has been a beam damage incident and if modules were exchanged. Mechanical position calibration is recommended every time the detector position is changed unless knowledge of the exact module positions is not required for the experiment. 

Finding a quicker way to reliably determine high quality calibration constants is an ongoing development effort together with the detector group from the European XFEL.

\section{Performance data and imaging example}

In order to test the noise performance over the entire dynamic range a pulsed IR laser was used, where the deposited energy in a pixel was varied between 1 and $10^4$ 12.4~keV photon equivalents by reducing the intensity of the IR pulses with calibrated filters. 

The noise of the system is higher in medium and low gain mode (approximately equivalent to 3.5~keV and 18~keV, respectively), but it was shown that it always remained significantly below the Poisson noise of the incoming signal \cite{mezza1, UT_IEEE14}. The same IR laser was used to scan the dynamic range of the system, which was measured to be 34.4~x~10$^6$ electrons, corresponding to approximately 10$^4$ photons of 12.4~keV. The non-linearity of the low gain state proved to be better than 0.5\% up to 5~x~10$^3$ photons \cite{UT_IEEE14}.

\begin{figure}
\includegraphics[width=\textwidth]{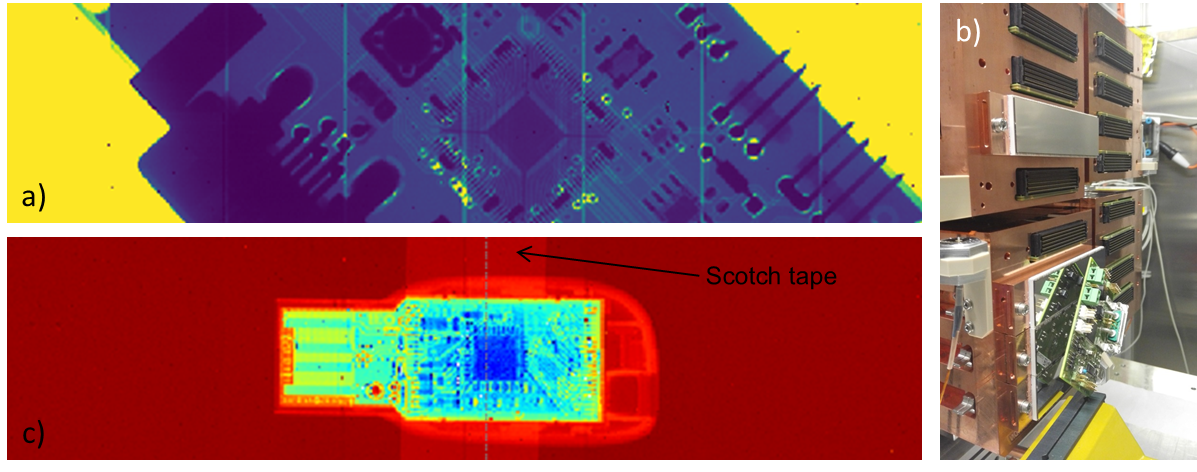}
\caption{a) Mean of 10,000 pedestal and gain corrected x-ray images of a PCB. b) AGIPD quadrants during mounting of front end modules. The imaged PCB is visible c) Mean of 30,000 x-ray images of a pen drive after pedestal, gain and flat field correction.}
\label{fig6}
\end{figure}

Fig. \ref{fig6}a shows the x-ray image of a printed circuit board taken with the setup shown in Fig. \ref{fig6}b. The shown x-ray image is the average of 10,000 individual images. 

Each individual image is corrected for pixel offset and gain on a per pixel basis. These corrections compensate fixed pattern ‘zero-level’ variations and pixel-to-pixel sensitivity variations that are commonly caused by process variations during the production of the ASIC, but can have many causes. 

Some artifacts remain after the correction. Especially in the medium intensity region vertical stripes can be observed. These originate in the double sized pixels (400~\textmu m~x~200~\textmu m) between ASICs. These pixels collect, on average, twice the amount photons, hence they appear brighter. 

Fig. \ref{fig6}c shows the x-ray image of a pen drive. For this image 30,000 individual images were averaged. 

Each image was offset and gain corrected as explained above, in addition a flat field correction was applied. The flat field correction accounts for the effective size of the pixels, removing the effect of the double size pixels, and removes artifacts from non-uniform illumination. Compensating the double sized pixels with a flat field correction is neither the only nor necessarily the best approach to correct for these pixels\footnote{For most experiments at the European XFEL these pixels are currently excluded from the data analysis.}. The image not only shows the structure of the plastic cover of the pen drive, but also the sticky tape which was used to hold the pen drive to the acrylic glass in front of the FEM, demonstrating the high sensitivity of the system. 

Lastly, the results of first user experiments show, that the detector is capable of determining protein crystal structures for both known and previously unknown proteins \cite{Ilme, SFX}. Of course this success is possible in combination with all the other infrastructure of the beamline at the European XFEL.

\section{Summary}

The adaptive gain integrating pixel detector, AGIPD, is an x-ray camera developed for use at the European XFEL. It was officially inaugurated together with the European XFEL in August 2017. 

This paper reviewed the complete system of the AGIPD 1 million pixel camera currently installed at the SPB beamline of the European XFEL. 

The system has a complex mechanical mounting that includes an in-vacuum movement system and many electronic boards, some of which are inside the vacuum vessel, some outside in an external housing. Its four independently movable quadrants can be arranged to form a horizontal slit or a rectangular hole with user selectable size.

The system is built from monolithic blocks of 2~x~8 ASICs forming a matrix of 128~x~512 pixels of (200~\textmu m)$^2$ size\footnote{If the double sized pixels are logically split the matrix is 128~x~526 pixels.}. Each pixel automatically adjusts to the incoming signal such that it can detect  any number of photons from single photons to 10$^4$ photons of 12.4~keV above its noise floor. Images are stored in one of 352 memory cells during the XFEL pulse train and read out in between trains.

The detector noise is approximately 1~keV (750~eV for high CDS gain), which is sufficient to detect single photons in many experiments and most of the early experiments at the SPB beamline have used the AGIPD with great success \cite{Ilme, SFX}.

The MID beamline of the European XFEL is scheduled to have a similar system installed in 2018.

\ack{Acknowledgements}

The authors are deeply indebted to the seemingly countless number of people that have contributed to this project. Beyond the former team members that are not included on the author list these are the people ‘behind the scenes’ at DESY, PSI, Hamburg University, Bonn University and the European XFEL, who, with their tireless efforts, created a joint environment that made the development of this camera possible.

We would like to explicitly thank the European XFEL detector group and the involved XFEL groups CAS and ITDM. Integrating a system as complex as the AGIPD into the environment at the European XFEL was an enormous undertaking that required a huge collaborative effort and was successfully done in time to operate the detector at the inauguration of the European XFEL facility.

We are thankful for the help from many of the DESY infrastructure groups (esp. ZM1, ZM2, ZM3, ZE, FEA and FEB) during the development of the AGIPD. 

We tested many prototypes, sometimes with help from external facilities. We would like to thank all of the people involved in these tests, which were crucial to the development process.

Last, but not least, we are very thankful for the valuable discussions with and the proof-reading of this manuscript by David Pennicard.





%
%


\end{document}